\documentclass[prb,twocolumn,superscriptaddress,amsmath,amssymb]{revtex4-1}
\usepackage{graphicx}
\usepackage{bm}
\usepackage{color}
\usepackage[normalem]{ulem}% provides \sout

\def\beq{\begin{equation}}
\def\eeq{\end{equation}}
\def\bea{\begin{eqnarray}}
\def\eea{\end{eqnarray}}
\def\fatR{{\bf R}}
\def\fatM{{\bf M}}
\def\fatr{{\bf r}}
\def\fatw{{\bm \omega}}
\def\fatM{{\bf M}}

\newcommand{\Ref}[1]{Ref.~\onlinecite{#1}}

\begin{document}

\title{
Fourier series of atomic radial distribution functions: 
A molecular fingerprint for machine learning models of quantum chemical properties
}
\author{O.~Anatole von Lilienfeld}
\email{anatole.vonlilienfeld@unibas.ch}
\affiliation{Institute of Physical Chemistry and National Center for Computational
Design and Discovery of Novel Materials, Department of Chemistry, University of Basel, Switzerland.}
\affiliation{Argonne Leadership Computing Facility, Argonne National Laboratory, 9700 S. Cass Avenue, Lemont, IL 60439, USA}
\author{Raghunathan Ramakrishnan}
\author{Matthias Rupp}
\affiliation{Institute of Physical Chemistry and National Center for Computational
Design and Discovery of Novel Materials, Department of Chemistry, University of Basel, Switzerland.}
\author{Aaron Knoll}
\affiliation{Mathematics and Computer Science Division, Argonne National Laboratory, Argonne, Illinois 60439, USA}
\affiliation{Texas Advanced Computing Center, University of Texas Austin, Texas, USA}

\begin{abstract}
We introduce a fingerprint representation of molecules
based on a Fourier series of atomic radial distribution functions. 
This fingerprint is unique (except for chirality), continuous, and 
differentiable with respect to atomic coordinates and nuclear charges.
It is invariant with respect to translation, rotation, and nuclear permutation,
and requires no pre-conceived knowledge about chemical bonding, topology, or 
electronic orbitals.
As such it meets many important criteria for a good molecular representation, 
suggesting its usefulness for machine learning models of molecular properties
trained  across chemical compound space.
To assess the performance of this new descriptor we have trained machine learning 
models of molecular enthalpies of atomization for training sets with up to 10\,k organic
molecules, drawn at random from a published set of 134\,k organic molecules
with an average atomization enthalpy of over 1770 kcal/mol.
We validate the descriptor on all remaining molecules of the 134\,k set. 
For a training set of 10k molecules the fingerprint descriptor
achieves a mean absolute error of 8.0 kcal/mol, respectively.
This is slightly worse than the performance attained using the Coulomb matrix,
another popular alternative, reaching 6.2 kcal/mol for the same training and test sets.
\end{abstract}

\maketitle

\section{Introduction}
For all but the most restricted problems and subsets of chemical compound space (CCS), screening, even when using high-throughput methods, becomes rapidly prohibitive due to the combinatorial explosion of possible arrangements of atom types and positions. 
The number of small stable organic molecules, for example, was estimated to exceed $10^{60}$.~\cite{ChemicalSpace}
The formal dimensionality of CCS corresponds to $4N$ degrees of freedom, 
associated to the three Cartesian coordinates and the one nuclear charge of $N$ atoms. This clearly illustrates the ``curse of dimensionality'' from which many first-principles inverse design efforts suffer~\cite{ZungerNature1999}.
Consequently, more compact representations of CCS are desirable,
in particular if they can be more intuitively dealt with.

Recent machine learning (ML) efforts leverage modern data analysis methods for atomistic simulations. 
The basic idea is to develop algorithms that {\em infer} the solution of the electronic structure problem for a new material, 
rather than investing in the computational time to numerically solve it,
with increasing accuracy as more training data are added~\cite{CurtaroloPRL2003}.
Given sufficient data, these approaches are among the 
most promising avenues towards efficient exploration of CCS from first principles. 
A more profound and rigorous understanding of CCS would greatly help 
computational design and optimization of new materials with desirable properties. 
As such, efficient navigation of CCS is at the heart of all first principles based materials and bio design efforts. 

To infer properties based on their correlations with compounds is
akin to what Hammett accomplished in the 1930s through
the exploitation of linear free energy relationships~\cite{Hammett_cr1935,Hammett_jacs1937}.  
Such approaches have already delivered convincing results for highly relevant applications, 
such as enhanced sampling~\cite{BEP4samplingGoedecker08}, 
screening of heterogeneous catalyst candidates based on Sabatier's principle~\cite{ReviewCatalystNorskov},  
and, devising simple materials design rules leading to topological insulators, semi-conductors, and others~\cite{Curtarolo2013}.

With increasingly available simulation data stemming from routine applications of
first principles methods, such as Born-Oppenheimer or Car-Parrinello molecular dynamics~\cite{AIMD_PNAS_TUCKERMAN2005},
statistical ML methods can be applied, in the hope of detecting trends 
and relationships that hitherto were difficult, 
if not impossible, to spot for the human expert. 
Applications of such approaches include data-mining for
crystal structure discovery~\cite{MachineLearningHautierCeder2010},
regression for reorganization energies that enter Marcus charge transfer rates~\cite{RatnerJACS2005,anatole-MilindDenis2011},
learning of potential energy surfaces (PES)~\cite{SumpterNoidNeuralNetworks1992,Neuralnetworks_Scheffler2004,NN_Tucker2006,Neuralnetworks_BehlerParrinello2007,bpkc2010}, 
learning of density functionals~\cite{ML4Kieron2012}, and
learning of the electronic Schr\"odinger equation of organic molecules~\cite{RuppPRL2012}.
The success of the latter, i.e., the success of predicting PESs across CCS without human bias yet in an accurate and reliable fashion, 
hinges on how well the input variables are represented for use by the ML algorithm.
This representation, also known as ``descriptor'', encodes chemical identity in terms 
of chemical composition and atomic configuration. 
As such, descriptors are a crucial ingredient for the
development of predictive ML models of PESs across CCS.

Conventionally, descriptors encode some prior knowledge about electronic structure effects. 
A frequently made assumption is that number and order of covalent bonds in compounds are known {\em a priori} and fixed.  
For example, Faulon's Signature descriptor encodes the graph of covalent bonding in a molecule~\cite{SignatureFaulon2003}.
Among others, this descriptor was applied to inverse QSAR~\cite{Visco2002}, 
prediction of protein interactions~\cite{FaulonProtein2005},  
and to predict reorganization energies of poly-aromatic hydrocarbons~\cite{anatole-MilindDenis2011}. 
The underlying assumption severely limits the realm of applications when it comes to modeling processes 
where the bonding is not known {\em a priori}. 
Examples for which this assumption is not valid include basically all processes commonly referred to
as ``chemical change'', i.e., bond breaking and formation, metal ligand exchanges, diffusion of defects in solids, 
proton hopping in aqueous solvents (Grotthuss-mechanism) or even simple tautomeric equilibria. 
The assumption also breaks down for interactions less localized than covalent binding, 
such as supramolecular van der Waals complexes, bound due to hydrogen bonds or (many-body) London dispersion forces~\cite{mbd_PNAS2012}.
The latter are particularly crucial for biological function, as recently illustrated for the selective self-assembly of hydrogen-bonded nano-structures~\cite{molecular-recog}.
For a comprehensive overview and comparative analysis of over 600 different descriptors
see the 2005 study carried out by Meringer and coworkers~\cite{DescriptroOverviewMeringer2005}.
The limitation due to such inherent assumptions might possibly explain 
the current state of affairs in QSAR-based drug discovery efforts~\cite{SchneiderReview2010}.
It is therefore desirable to devise more general ``first principles-like'' descriptors 
that conserve the rigor of {\em ab initio} methods~\cite{AbInitioDefinitionByKieronBurke}, 
such as wave function~\cite{MolecularElectronicStructureTheory}, 
density functional~\cite{TruhlarDFTreview2008}, or quantum Monte Carlo
methods~\cite{QMC_HydrogenOngraphene_AngelosDario2011},  
and that consequently can also, at least in principle, 
account for {\em any} chemical scenario or compound~\cite{ChemicalSpace,anatole-jcp2006-2}. 

In this study we introduce a molecular descriptor that uniquely (except for chirality) 
represents any molecule as a fingerprint, here a univariate function in terms of geometric distance.
Within the Born-Oppenheimer view on the CCS of molecules, 
{\em any} molecular geometry is uniquely characterized within its 4$N$-6 degrees of freedom, 
subtracting three rotational (two if linear) and three translational degrees of freedom.
Our descriptor is unique, differentiable, and invariant with respect to rotation, translation, and indexing of atoms. 
In full analogy to the information entering the electronic Schr\"odinger equation, 
this descriptor requires {\em only} atomic coordinates and nuclear identities. 
Thus, any composition and geometry is accounted for in a way amenable to ML. 
The descriptor might even be suitable for modeling of nuclear quantum effects through {\em ab initio} path-integral 
molecular dynamics~\cite{tuckerman_book_SM}, relevant for instance in the case of 
of Watson-Crick tautomers~\cite{anatole-jacs2010}, 
if energies and forces for all the replicas can be learned with sufficient predictive accuracy. 

This paper is structured as follows. 
In the methods section, we first outline the conceptual framework 
and discuss desirable properties of descriptors. 
Then, starting with the external potential, we proceed with a step by step 
discussion of translational, rotational and atom-indexing invariances,
as well as uniqueness requirements, which have guided us to the specific form of our descriptor. 
In the results section, the descriptor's performance is assessed and compared 
to the Coulomb matrix, another popular descriptor, 
using heats of atomization of up to 134\,k organic molecules taken from \Ref{DataPaper2014}.

%But also other applications, such as structural search methods, rely on measuring distances
%in configurational space. 
%For example, the minima hopping by Goedecker et al uses ...
%Cite
%Metrics for measuring distances in configuration spaces, Ali Sadeghi, S. Alireza Ghasemi, Markus A. Lill, Stefan Goedecker
%Or the Learn-On-The-Fly uses Gaussian atomic density overlap integrals.
%Cite Sandro.

%What about Wilson's B-matrix????

\section{Method}
\subsection{Descriptor properties}
The defining purpose of a descriptor $D$ is to represent a compound, 
defined through input variables, in a form that can be correlated to a property of interest~$\mathcal{P}$,
i.e., its form should be amenable to statistical learning. 
More specifically, $D$ should rigorously {\em and} in a convenient fashion represent the variables
that occur in the equation being modeled via ML.

Many descriptors and classification schemes for them have been proposed.
For the purpose of modeling results derived from Schr\"odinger's equation, 
one could consider the following three cases
\begin{description}
\item[First principles] Descriptors that encode the relevant information in the quantum Hamiltonian without loss of information. 
As such, they should be applicable to the learning of any quantum observable, such as energies, forces, or electronic properties. 
Examples include the sorted Coulomb-matrix~\cite{RuppPRL2012}, Gaussian shapes~\cite{GaussianShapeDescriptor1996}, 
bispectrum, power spectrum, or angular distribution functions~\cite{bpkc2010,BartokGabor_Descriptors2013}. 
The challenge consists of removing redundancies and encoding invariances, i.e., to render them maximally compact without losing information. 
Note however, that some observables might require more degrees of freedom than others. 
In ML models of atomization energies, for example, the chirality of the molecule is not relevant. 
For ML models of the optical activity in circular dichroism, however, it is. 

\item[Coarsened] Descriptors that reflect important structural features typically work 
for a range of properties but not for all of them. 
Examples include the number of hydrogen-bond donors or acceptors (used in Lipinski's rule of five\cite{lldf1997}), 
number of aromatic units, the diagonalized Coulomb matrix~\cite{MoussaComment,MoussaReply}, 
the bag-of-bond descriptor~\cite{BobPaper},
or the signature descriptors~\cite{SignatureFaulon2003}.
Such descriptors are not bijective, i.e., they do not allow reconstruction of the compound in general; 
in practice, some allow reconstruction given enough constraints.
The challenge consists of finding a form for which the loss of information is minimal while maintaining the advantages of coarsening.

\item[Integrated] Descriptors that explicitly encode integrated properties correlating well with the property of interest. 
Examples are adsorption energies for catalytic activity~\cite{ReviewCatalystNorskov}, logP octanol/water partition coefficients or Lipinski's rule of 5~\cite{ChemicalSpace} for oral bio-availability, electrophilic superdelocalizability for p$K_a$ prediction\cite{tlwpmmg2002}, 
HOMO eigenvalues~\cite{NN4B3LYP_Chen2003,NN4B3LYP_Chen2004} and other simple property descriptors commonly used in high-throughput screening~\cite{Curtarolo2013}.
The challenge consists of gauging their transferability to other compound classes, properties, or even environmental conditions.
\end{description}
In this study, we restrict ourselves to {\bf First principles} kind of descriptors which can be used for
the construction of ML models of quantum mechanical observables.
For {\bf Coarsened} or {\bf Integrated} descriptors the reader is referred to the above cited literature.

\subsubsection*{Uniqueness}
We believe uniqueness, up to invariants that leave the modeled observable $\mathcal{O}$ unchanged, to be crucial. 
In other words, we consider a descriptor to be unique if there is no pair of  molecules 
that produces the {\em same} descriptor. 
Here, we do not refer to the reverse case, namely that any given 
{\em single} molecule can have more than one descriptor.
For example, our criterion of uniqueness is still met by any descriptor consisting simply of the set of atomic nuclei and associated Cartesian coordinates 
due to the mapping between molecular Hamiltonian and unperturbed wave-function $\Psi$ of the ground-state:
While no pair of molecules exists with the exact same sets of atomic nuclei and coordinates,
there are many different sets of coordinates which merely differ by  molecular symmetry operations 
(translation, rotation, or complete nuclear permutation)~\cite{BunkerSymmetry}.
Removal (or reduction) of such invariant degrees of freedom
%---apart from translations and rotations there should also be permutational invariance for all atoms with the same nuclear charge---
is relevant for the efficiency of the machine learning model
but less crucial on a conceptual level.

The reason for the uniqueness requirement can be shown by {\em reductio ad absurdum} in three 
steps---in analogy to the first Hohenberg-Kohn theorem~\cite{HK}---for 
any quantum mechanical observable $\mathcal{O} = \langle \Psi | \hat{O} |\Psi \rangle$.
Here the unperturbed ground-state Hamiltonian $H$ is defined by its external potential, determined by
$\{Z_I, \fatR_I\}$, the set of nuclear charges and coordinates, as well as number of electrons $N_e$.
The variational principle yields the system's many-body wavefunction $\Psi$ for any given $H$.
\begin{itemize}
\item[(i)] Let $D$ denote a descriptor that is {\em not} unique. %up to all relevant invariant degrees of freedom.
Then two systems $H_1 \neq H_2$ exist that differ in excess of the invariants, 
but they are mapped to the same descriptor value $d$, $H_1 \rightarrow d$ and $H_2 \rightarrow d$.
\item[(ii)] Because $H_1$ and $H_2$ differ by more than their property's invariances, 
they have different wave-functions, $\Psi_1 \neq \Psi_2$, 
yielding two different observables, 
$\mathcal{O}_1 = \langle \Psi_1 | \hat{O} |\Psi_1 \rangle$
and 
$\mathcal{O}_2 = \langle \Psi_2 | \hat{O} |\Psi_2 \rangle$.
Here, we deliberately ingore the obvious exception and special situation of all observables which happen to be degenerate. 
\item[(iii)]
A trained statistical model predicts any observable $\mathcal{O}$ solely based on descriptor input $d$
leading to identical predictions $\mathcal{O}_1^{\mbox{\scriptsize pred}} = \mathcal{O}_2^{\mbox{\scriptsize pred}}$. 
In the limit of infinite training data, these predictions will be exact,
implying $\mathcal{O}_1 = \mathcal{O}_2$, in contradiction to (ii).
\end{itemize}
Consequently, non-unique descriptors can yield absurd results for {\em any} observable. 
In other words, artificial degeneracies in the descriptor imply prediction errors that can not be 
remedied through addition of more training data. 
As such, non-unique descriptors defy the very idea of using ML in quantum mechanics.

Uniqueness up to invariances is necessary, but not sufficient for the design of a good descriptor.
Consider the case of the invariant degrees of freedom, 
for which a manifold of unique descriptors could be constructed:
Descriptors which depend on rotations, translations, and nuclear permutations could in principle be used,
the obvious example being the $4N$-dimensional vector with 
four entries corresponding to nuclear charge and three Cartesian coordinates, $Zxyz$.
For example, translational invariance could be imposed by including shifted copies of $Zxyz$ vectors in the training set. 
While representations of internal degrees of freedom, such as atom-atom distance matrices or 
the Z-matrix, popular in quantum chemistry communities, encode rotational and translational
invariances, they still suffer from lack of nuclear permutational invariance.
It is possible to obtain invariance with respect to nuclear permutation by 
simply representing each molecule not by one but rather by a set of $Zxyz$ vectors, 
each vector containing the same elements but in different order (see~\Ref{Montavon2013} for a successful application of this idea).
However, in general such descriptors lead to substantial overhead for the statistical learning, since in order to
obtain a transferable ML model, the training set would have to be constructed (and extended) to explicitly reflect all these invariances.
%Such ``solutions'' would lead not only to inefficient ML models, 
%but also to training set sizes growing factorially with number of atoms. 
Furthermore, the model's transferability would also be inherently limited to those ranges of the redundant degrees 
of freedom that have been covered in training.
Also, when it comes to measuring similarity between descriptors of two molecules (the ultimate feature of any ML model)
absence of translational, rotational, and nuclear permutation invariance can aggravate alignment problems,
with multiple minima, and numerically difficult and challenging optimization problems,
as recently reviewed by Zadeh and Ayers~\cite{AlignmentAyers2013}.
Another reason for aiming to remove all invariances can be given by analogy to the definition of the property. 
In the case of the electronic Schr{\"o}dinger equation, position and orientation of the external potential in the Hamiltonian 
are arbitrary, and the external potential, a sum over all nuclei, is permutationally invariant.
Since the descriptor is meant to represent the independent variables in the Hamiltonian, 
it suggests itself that it be invariant with respect to all the redundant degrees of freedom. 
Finally, absence of invariance with respect to nuclear permutations might also become cumbersome for modeling the energy when it comes 
to simulation regimes in which Heisenberg's uncertainty principle applies to atoms, such as collisions at high temperature, and
when atoms of the same type and weight can become indistinguishable.
We conclude that a lack of invariances can present severe challenges in practice, 
it appears therefore desirable to map all invariant structures to the same descriptor, i.e., for the descriptor to obey all invariances.
The challenge consists of removing as many of these redundant degrees of freedom as possible, {\em without} losing uniqueness, 
i.e.~without turning the {\bf First principles} descriptor into a {\bf Coarsened} descriptor.
Below we will encounter an example for a coarsened descriptor ($FD$) that meets all invariances but has lost uniqueness 
[see Eq.~(\ref{eq:FD_1}) and Fig.~(\ref{fig:homometric})].

We reiterate that the uniqueness vs.~invariance issue is strongly dependent on property. 
For example, atomization energies of stereo-isomers, calculated within non-relativistic 
Born-Oppenheimer-approximated time-independent electronic structure theory,
do not violate parity and are therefore also invariant with respect to choice of enantiomers~\cite{QuacksParity}.
If the goal was to also account for parity-violating effects in the potential energy, 
the descriptor would have to be enantio-selective. 

\subsubsection*{Desirable properties}
The descriptor's size-extensive and symmetry behavior is also relevant. 
In analogy to the external potential in Schr\"odinger's equation, 
atoms or groups that are symmetric should contribute in equal ways to the descriptor;
and changes in system size should lead to corresponding changes in the descriptor (e.g., its size or range of component values).
Another important feature of a descriptor is its completeness, or global nature, meaning that it encodes the whole of a compound, as opposed to only a local part of it.
Local descriptors in terms of expansions over atomic contributions were successfully used 
with neural networks~\cite{Neuralnetworks_BehlerReview2011} and kernel ridge regression~\cite{bpkc2010,BartokGabor_Descriptors2013} to learn potential energy hypersurfaces and forces across configuration space, 
and, for enhanced sampling using molecular dynamics. 
While local descriptions form the basis for linear-scaling electronic structure software,
and might be appropriate for some properties (e.g., properties of atoms in molecules, such as nuclear magnetic resonance chemical shifts or atomic forces) and systems (e.g., insulators and semi-conductors where the electronic ``nearsightedness principle''~\cite{KohnNearsightedness} can be exploited), 
this can not be assumed in general, and might be limiting for energies of long-range electron/phonon coupling, electron transfer, or metals. 

Other desirable features of descriptors include 
a closed and analytic form for analysis and rapid evaluation, 
differentiability (with respect to nuclear charges and coordinates) to account for response properties and use of advanced learning techniques, 
uniform length for finite sets of compounds to conveniently compare molecules that differ strongly in size (number of atoms), 
and, a functional form that can cope well with all the various ranges relevant to physical chemistry, 
i.e.~nuclear charges ranging from 0 to $\sim$100,  
and interatomic distances ranging from tenths to dozens of~{\AA} 
(or even a thousand~{\AA} in the case of more exotic molecules~\cite{LongBondRb2}). 

\begin{table}
\caption{
Properties of various descriptors including Signature ($\sigma$) ~\cite{SignatureFaulon2003}, nuclear charges and Cartesian coordinates ($Zxyz$),
Coulomb matrix ($CM$), diagonalized $CM$ matrix (Eig($CM$))~\cite{RuppPRL2012}, and the radial distribution Fourier series descriptor ($FR$) introduced here.
$N$ denotes number of atoms. 
The upper part contains requirements used for the design of FR.
$\checkmark$ and $\neg$ indicate whether a requirement is fulfilled or not.
}
\label{tab:Properties}
\begin{tabular}{|l|c|c|c|c|c|} \hline 
Property                 & $\sigma$  & $Zxyz$     & $CM$        & Eig($CM$) & $FR$     \\\hline
Unique                   & $\checkmark$   & $\checkmark$   & $\checkmark$   & $\neg$ & $\checkmark$ \\ 
First principles         & $\neg$    & $\checkmark$  & $\checkmark$   & $\checkmark$ & $\checkmark$ \\ 
Transl. invariant        & $\checkmark$   & $\neg$   & $\checkmark$   & $\checkmark$ & $\checkmark$ \\ 
Rotat. invariant         & $\checkmark$   & $\neg$   & $\checkmark$   & $\checkmark$ & $\checkmark$ \\ 
Index. invariant      & $\neg^a$    & $\neg$   & $\neg$    & $\checkmark$ & $\checkmark$ \\ 
Differentiable           & $\neg$.   & N.A.     & $\checkmark$   & $\checkmark$ & $\checkmark$ \\ \hline
Symmetry                 & $\checkmark$   & $\neg$   & $\checkmark$   & $\checkmark$ & $\checkmark$ \\ 
Size extensive           & $\checkmark$   & $\checkmark$  & $\checkmark$   & $\checkmark$ & $\checkmark$ \\
Complete/global          & $\neg$~\footnote{unless taken to full height $h=N$.}   & $\checkmark$  & $\checkmark$   & $\neg$  & $\checkmark$ \\  
Dimensionality           & N.A. & $4N$     &$(N^2-N)/2$&    $N$  & $m$~\footnote{$m \ge (N^2-N)/2$ being the number of grid elements required for discretizing the largest interatomic distance} \\ 
Analytical               & $\checkmark$   & $\checkmark$  & $\checkmark$   & $\checkmark$ & $\checkmark$ \\ 
Uniform length           & $\neg$    & $\neg$   & $\neg$    & $\neg$  & $\checkmark$~\footnote{If damped by a Gaussian} \\ 
Variable ranges          & $\checkmark$   & $\checkmark$  & $\checkmark$   & $\checkmark$ & $\checkmark$ \\  \hline
\end{tabular}
\end{table}
%length invariance /size extensive
%differentiable wrt all Zxyz

An overview of crucial and desirable properties is given in Table~\ref{tab:Properties} for various relevant descriptors,
including Signature ($\sigma$)~\cite{SignatureFaulon2003}, nuclear charges and Cartesian coordinates ($Zxyz$),
Coulomb matrix ($CM$), diagonalized $CM$ matrix (Eig($CM$))~\cite{RuppPRL2012}, 
and the Fourier series of atomic radial distribution functions ($FR$), introduced here.
$CM$, recently introduced~\cite{RuppPRL2012}, satisfies many of the aforementioned requirements, but not all.
In particular, it lacks invariance with respect to nuclear permutation 
(fixed in practice by sorting atom indices with respect to the norm of it's rows or columns),
%it's not uniform in length, 
and its dimensionality scales quadratically with number of atoms. 
We note that the dimensionality listed in Table~\ref{tab:Properties}, however, is a rather formal construct: 
The $(N^2-N)/2$ entries in the Coulomb matrix are not independent variables.
%as such redundant dimensions have been introduced which do not change the underlying dimensionality of the problem.
This statement is clearly also true for the $m$ grid-points representing ``dimensions'' of the $FR$ descriptor.

Another aspect is the smoothness of the property as a function of the descriptor.
Smoothness is a prerequisite for machine learning (to enable meaningful selection from the infinitely many models that are compatible with the training data), related to regularization.
However, the function's smoothness might vary along different directions in descriptor space
(as an example, consider ligand binding, where steric constraints of the host might cause abrupt changes in affinity upon certain geometrical changes of the ligand, as opposed to more gradual changes not conflicting with the host's geometry, e.g., ``magic methyls'', or, more generally, ``activity cliffs'' \cite{m2006d}).
Reducing the models smoothness requires more training data than necessary in smooth data regions, whereas increasing the models smoothness reduces prediction accuracy in data regions with more rapid changes.
A potential solution might be models with local smoothness ~\cite{pkb2008}.

In the following we discuss the sequence of steps that has led us to the $FR$ descriptor which meets all the required and desired features listed in Table~\ref{tab:Properties}, 
i.e.~(i) first principles and nuclear permutation invariance, (ii) translational invariance, (iii) rotational invariance and mirror symmetries (Euclidean symmetries), (iv) uniqueness, and (v) differentiability.

\subsection{First principles {\em Ansatz}: The external potential}
The first Hohenberg-Kohn theorem shows that the electron density $n(\fatr)$ of a given system, as determined by its external potential $v(\fatr)$ through application of the variational principle,
is as unique as its electronic wavefunction $\Psi(\fatr)$ obtained through solution of Schr\"odinger's equation~\cite{HK}. 
After application of the variational principle (yielding the electron density that minimizes the energy), 
the total potential energy is commonly given as an integral containing electron density and external potential,
\bea
E[n(\fatr)] & = & F_{ee}[n(\fatr)] - \int d\fatr \; n(\fatr) v(\fatr) + \frac{1}{2} \sum_{IJ} \frac{Z_IZ_J}{|\fatR_I - \fatR_J|}, \nonumber \\
\label{eq:DFT}
\eea
with $F_{ee}$ the universal functional encoding all contributions to energy coming from electron-electron interactions,
the second term representing the Coulomb attraction between nuclear charges and electrons, and the last term corresponding to the nuclear Coulomb repulsion between all atoms.
The total potential energy of {\em any} molecule is therefore determined, independent of translations, rotations, or nuclear permutations, determined by its unique electron density.

The electron density can be viewed as a ``quantum'' molecular descriptor, used to predict molecular energies through the map $n(\fatr) \mapsto E$. 
Already three decades ago, Carb\'o et al. proposed to use the overlap integral of electron densities of different molecules to 
quantify molecular similarity.~\cite{SimilarityCarbo1980}
In fact, the electron density is already used as a descriptor in practice when density functionals are trained empirically to reproduce the energies of a training set.
If the electron density were not unique, density functional theory as we know it would not exist. 
The Hohenberg-Kohn theorem in this sense underscores the importance of the descriptor's uniqueness when it comes to the training of potential energy surface models.

The external potential, conversely, is in a unique relationship with atomic Cartesian coordinates $\{\fatR_I\}$ and
nuclear charges $\{Z_I\}$, $v(\fatr) = \sum_I Z_I/|\fatr-\fatR_I|$.
%Note that the divergence at $\fatr = \fatR_I$ can be switched off through use of a soft Coulomb potential, $Z_I/(|\fatr-\fatR_I| + b)$,
%with $b > 0$, yielding a finite value of $Z_I$ for $v(\fatr=\fatR_I)$.
%As such, any molecule is represented simply by its electron number and its distribution of classical nuclear point charges in real space,
%$\sum_I^N Z_I \delta{|\fatR_I-\fatr|}$~\cite{anatole-jcp2006-2}.  
Due to its translational and rotational dependence, however, the external potential
itself does not qualify as a promising descriptor.
In a first step we replace the system's representation in form of its external potential 
by a model of nuclear charge densities, namely a sum of
Gaussians located at atomic coordinates with atom type-specific heights $Z_I$, 
\bea 
P(\fatr) & = & \sum_I  Z_I e^{-a|\fatr-\fatR_I|^2} ,
\label{eq:ZofR} 
\eea 
where the sum runs over all $N$ atoms in the molecule, and $a$ is a global parameter for all atoms and all molecules, for now simply fixing the nuclear width for {\em all} atoms independent of type. 
Note that $a$ could be defined in an atom-type specific way, and that $P$ does no longer integrate to $N_p$, the total number of protons present in the system, except for infinitely small width of the Gaussians.
Similar to Gaussian type orbitals as a basis for molecular orbitals, 
we thereby deliberately forego any physical non-Gaussian like features in favor of computational convenience. 
Note, however, that $P(\fatr)$ is still in a one-to-one relationship with the external potential, 
and that it is still atom index invariant.

\subsection{3-D Fourier transform}
%Fourier transforms in spectroscopy and plane wave basis sets.
An appealing characteristic of using plane wave basis sets in electronic 
structure calculations is their invariance with respect to atomic position (translational invariance).
In contrast to atomic basis sets, Pulay forces and basis set superposition errors,
i.e., additional force terms due to basis set incompleteness, can be avoided, 
which makes the implementation of geometry optimization or molecular dynamics methods more straight-forward.
Analogously, we can remove translational degrees of freedom of a charge distribution by changing 
to the Fourier frequency domain~\cite{RotationalInvarianceFourier2009}.
The Fourier transform of the Gaussian charge distribution, 
\bea
\mathcal{F}(P) & = & \frac{1}{(2a)^{3/2}} e^{-\frac{\fatw^2}{4a}} \sum_I Z_I e^{i\fatw^T\fatR_I},
\eea
can be multiplied with its conjugate to yield a real function in the three dimensions of the Fourier domain $\fatw$,
\bea
F(\fatw) & = &
 \frac{1}{(2a)^{3}} e^{-\frac{\fatw^2}{2a}} \sum_{J,I}  Z_IZ_J \cos [\fatw^T(\fatR_I-\fatR_J)],
\label{eq:FT}
\eea
after simplification using Euler's formula. 
Eq.~(\ref{eq:FT}) is a translation-invariant representation of the nuclear charge distribution in Eq.~(\ref{eq:ZofR}).
$F(\fatw)$ can be viewed as a sum over all elements of a 
symmetric atom-atom pairwise matrix $\fatM$ with elements
\bea
M_{IJ} & = & Z_I Z_J \cos[\fatw^T(\fatR_I - \fatR_J)].
\eea
This matrix is reminiscent of the Coulomb matrix~\cite{RuppPRL2012}. 
At $\omega = 0$ and for $a = (1/4)^{1/3}$, its diagonal elements become identical
to those of a preliminary version of the Coulomb matrix, 0.5 $Z_I^2$, the potential
energy of the hydrogenic atom.
While Eq.~(\ref{eq:FT}) has appealing features, it still lacks rotational invariance. 
Furthermore, preliminary tests with machine learning models of atomization energies based on this descriptor,
after alignment of all molecule pairs in the Fourier domain, resulted in rather disappointing predictive accuracy for out-of-sample molecules.

\subsection{1-D version}

\begin{figure*}
\centering
\includegraphics[scale=0.23, angle=0]{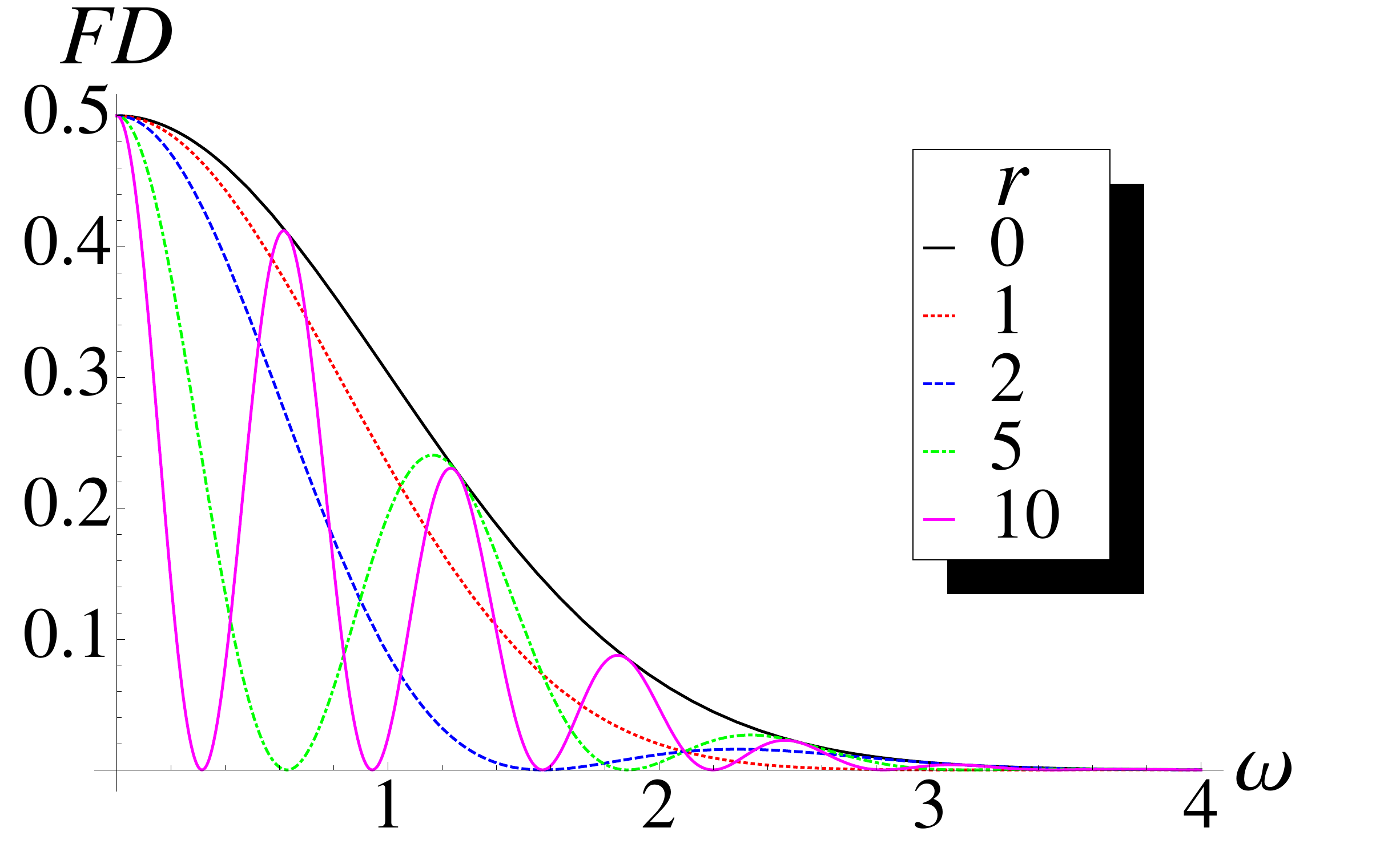}
\includegraphics[scale=0.23, angle=0]{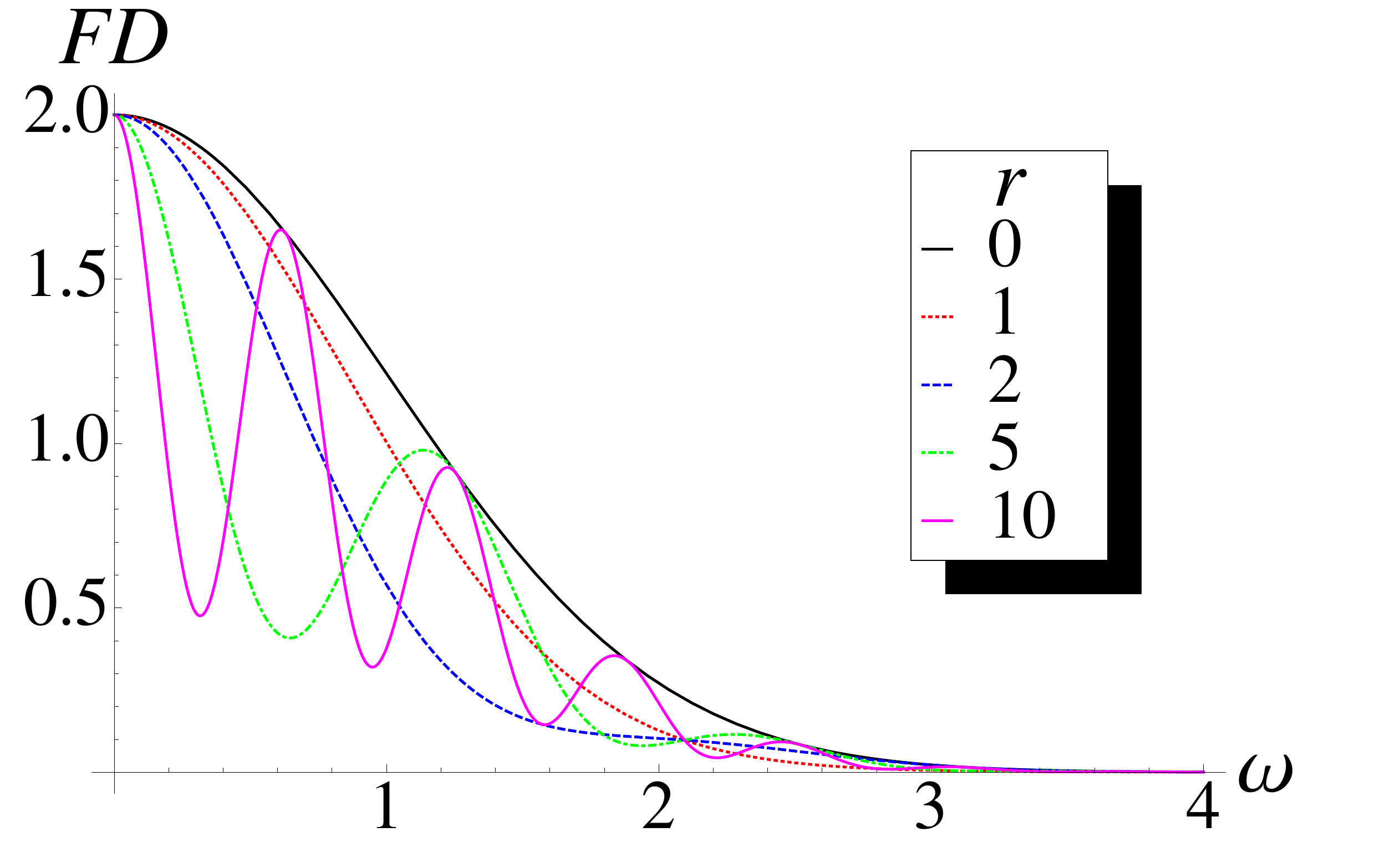}
\includegraphics[scale=0.23, angle=0]{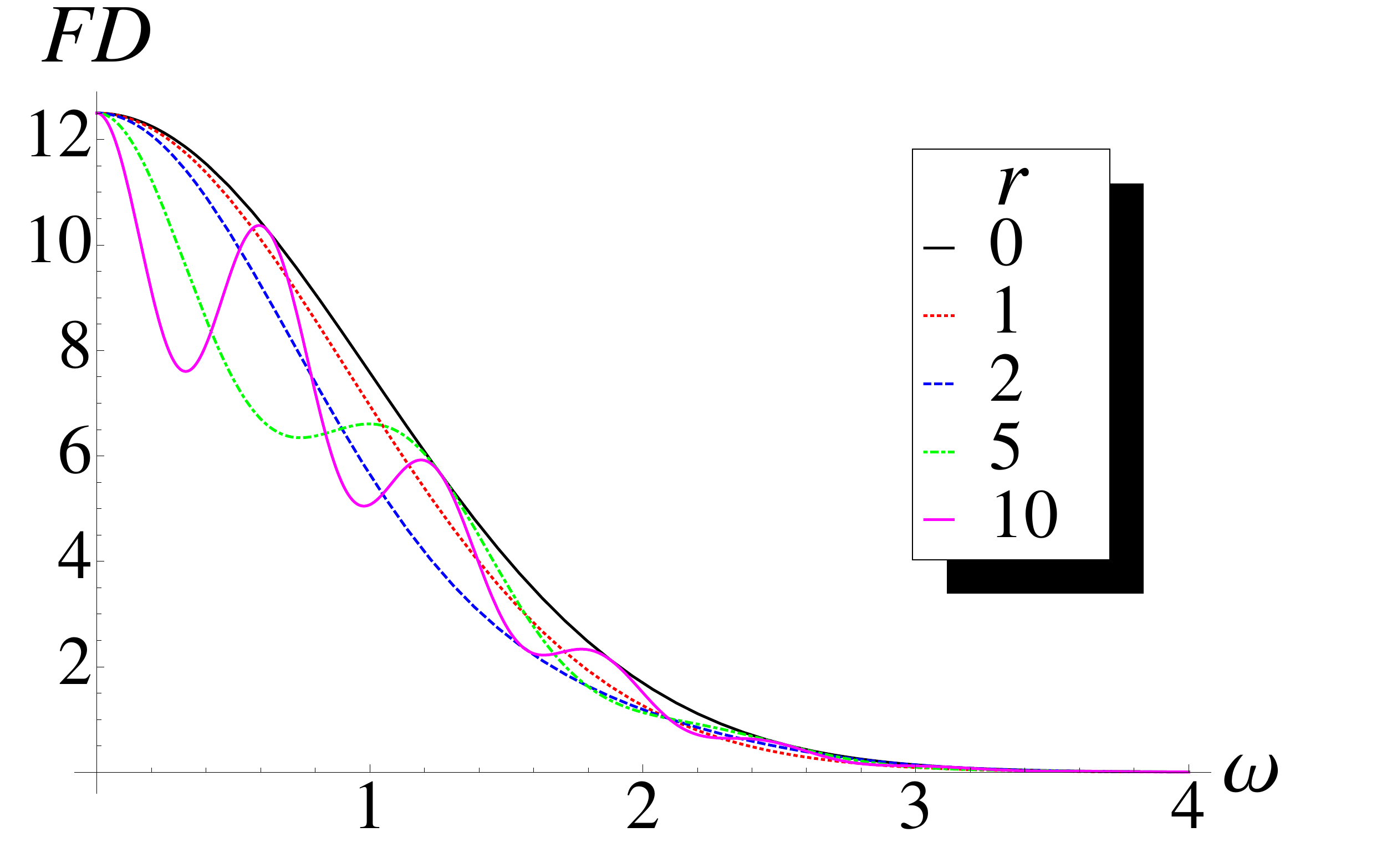}
\caption{
Illustration of Fourier descriptor $FD$ (units of charge squared, Eq.~(\ref{eq:FD_1})) for three diatomics, H$_2$ (left), LiH (center), and HF (right) 
for five interatomic distances $r$. 
The hyperparameter $a$ is set to 1.
Note that for $r$ = 0, $FD$ corresponds to $(Z_1^2+2Z_1Z_2+Z_2^2)/8$.
}
\label{fig:FD_1}
\end{figure*}

To remove rotational dependence, we project the Fourier transform (Eq.~\ref{eq:FT}) onto one dimension by replacing the argument of the cosine function by the scalar product of a frequency and the interatomic distance:
%by $\fatw^T(\fatR_I - \fatR_J) \mapsto \omega \times |\fatR_I - \fatR_J|$,
% f7 = \sum_I Z_I^n_(1,Z)( cos[Z_I^n_(2,Z) \sum_J Z_J^n_(3,Z) * exp()] -  1) 
% f7 is variational in b and n_(i,Z). Here i \
\bea
\label{eq:FD_1}
FD(r) & = & \frac{1}{(2a)^{3}} e^{-\frac{\fatw^2}{2a}} \sum_{J,I} Z_IZ_J \cos[\omega \times r_{IJ}],
%\sum_I Z_I^{n_1}\left(\cos\left[\sum_J \frac{Z_I^{n_2}}{Z_J^{n_3}} e^{-b (r_{ij}-r)^2}\right] -  1\right)  \nonumber\\
\eea
where $r_{IJ} = |\fatR_I - \fatR_J|$ is interatomic distance.
Eq.~\ref{eq:FT} is a double sum over atoms that maintains invariance with respect to nuclear permutation, translation, and rotation.
%and where $\{n_i\},b$ are variational global hyper parameters which are optimized.
Fig.~\ref{fig:FD_1} 
%and ~\ref{fig:FD_2} 
illustrates the meaning of this "hack" for various interatomic distances of H$_2$, LiH, and HF.
%From Eq.~(\ref{eq:FD_1}), %\ref{eq:FD_2}),
Changes in interatomic distances induce changes in oscillatory frequency,
while changes in elemental composition affect overall amplitude. 
Differences in atomic numbers within any of these three diatomics 
are expressed through the width of the oscillatory band, i.e., the larger the difference, the narrower the band.
The Gaussian prefactor dampens the descriptor towards zero for large frequencies.

%\begin{figure*}[htbp!]

\begin{figure*}%{r}{13cm} %\vspace*{-0.0cm}
\includegraphics[scale=0.53, angle=0]{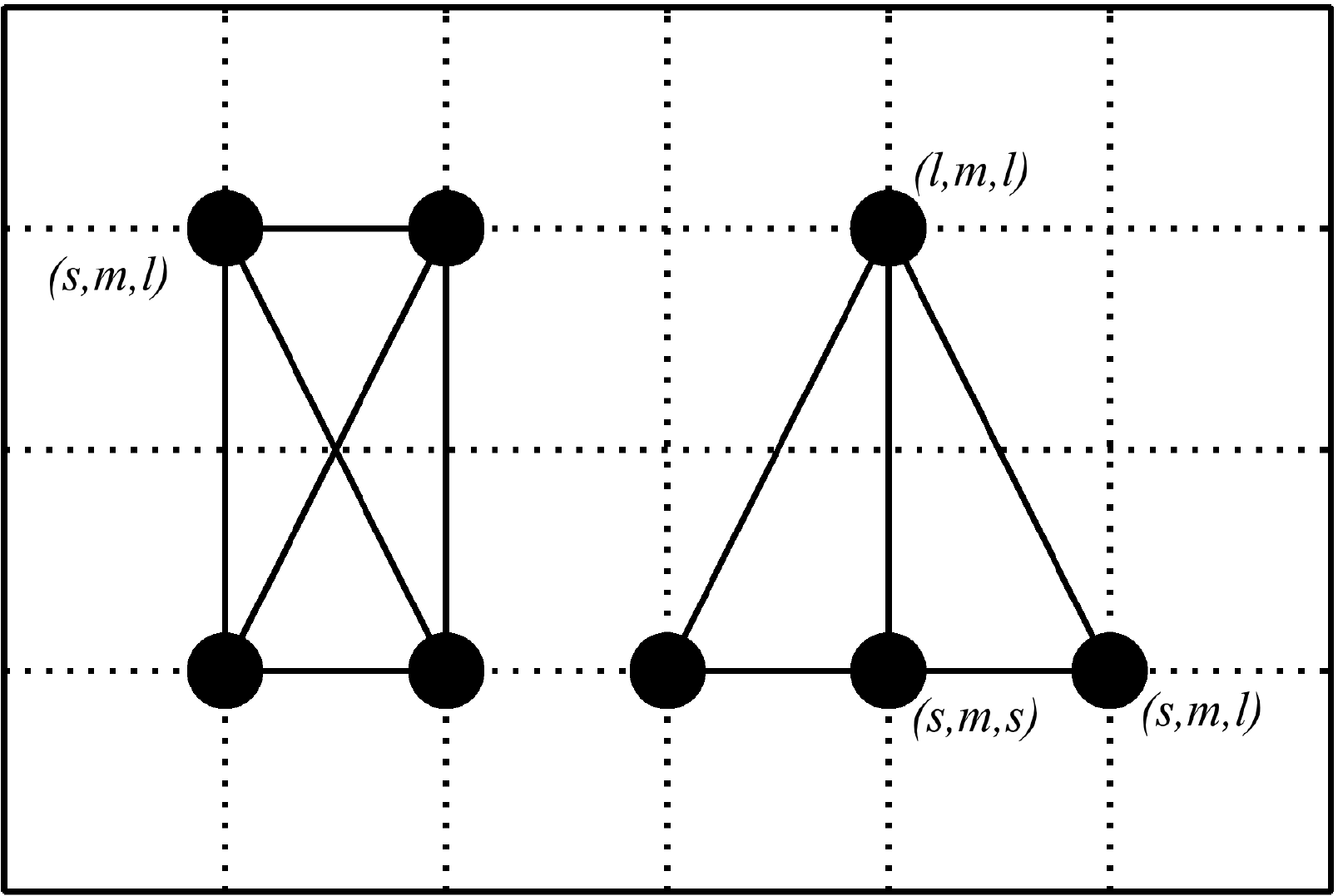}
\includegraphics[scale=0.34, angle=0]{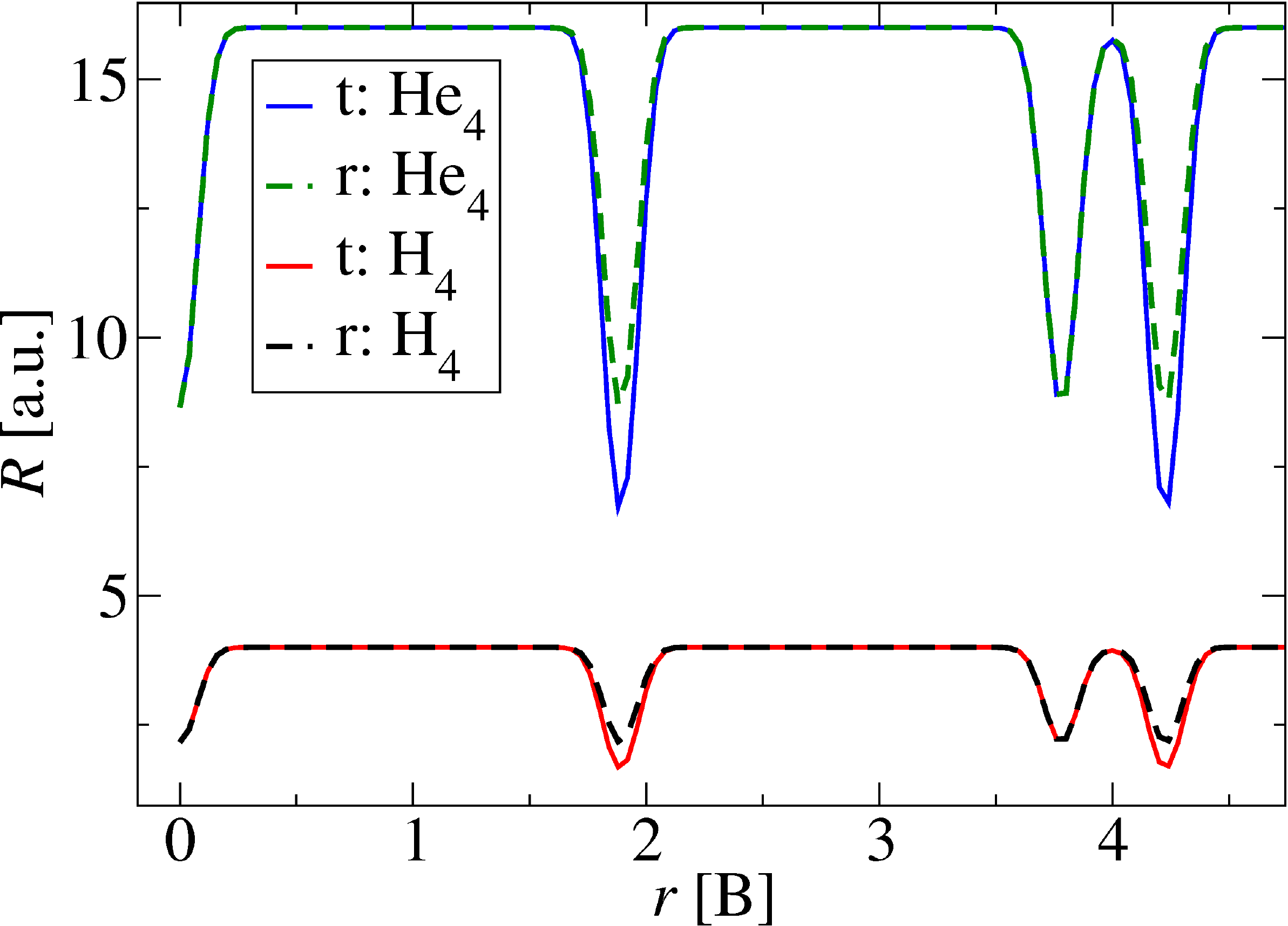}
\caption{
\label{fig:homometric}
\label{fig:He4andH24}
LEFT: Sketch of two homometric molecules (same atom types, same sum of interatomic distances) 
from \Ref{TruhlarHomometric}.
The atomic distribution of distances ($s$ short, $m$ medium, $l$ long) are indicated.
The sorted Coulomb matrix can distinguish these two molecules~\cite{NatureHomometric,MoussaComment,MoussaReply}.
$s$ and $l$ are set to 1 and 2 {\AA}, respectively.
%Sorted is not differentiable.  
%Permutations disadvantageous: Because of the poor scaling it might prove problematic, however,
RIGHT: Illustration of Fourier series with Gaussian radial 
distribution function based descriptor $R$ (according to Eq.~(\ref{eq:FDR}))
for the homometric, rectangular (r) and triangular (t), 
geometries displayed in Fig.~\ref{fig:homometric}. %and $b$ = 0.1.
}
\end{figure*}

\subsection{Uniqueness}
The 1-D Fourier fingerprint (Eq.~(\ref{eq:FD_1})) is invariant with respect to translations, rotations, and nuclear permutations.
However, it is no longer unique due to the information lost in modifying the argument of the cosine.
This can easily be seen for the task of distinguishing homometric molecules, i.e., molecules with identical sets of interatomic distances.~\cite{NatureHomometric} % between all the same atom-pairs.
Note that while in \Ref{MoussaReply} it is mentioned that all enantiomers are homometric, 
there exist also homometric compounds that are not enantiomers. 
While potentially of interest for the ML modeling of parity violation~\cite{QuacksParity}, 
for the electronic Schr{\"o}dinger equation within the Born-Oppenheimer approximation any mirror symmetries 
(leading to enantiomers) represent only redundant degrees of freedom, 
which need not be distinguished by the descriptor.
However, all pairs of homometric molecules that are not enantiomers should be distinguished by the descriptor.
An example of such a compound pair, proposed in \Ref{TruhlarHomometric}, 
is on display in Fig.~\ref{fig:homometric}. 
Note that any two homometric compounds would have exactly the same potential energy if modeled by an exclusively pair-wise interatomic potential,
no matter how well parametrized to effectively account for many-body effects. 
As such homometric compound pairs exemplify the importance of many-body effects in interatomic potentials, 
effects recently shown to be sizeable not only for covalent bonding, but also
for intermolecular van der Waals forces~\cite{anatole-jcp2010,mbd_PNAS2012}.

%\begin{figure}%{r}{13cm} %\vspace*{-0.0cm}
%\includegraphics[scale=0.35, angle=0]{H2_Gaussian.ps}
%\caption{
%\label{fig:Width}
%Effect of width $b$ on descriptor for H$_2$ at 2 {\AA} interatomic distance,
%[See Eq.~(\ref{eq:FR})].
%}
%\end{figure}

Homometric compounds do not differ by the number of ``bonds'' (interatomic distances), but rather by their distribution: 
In the rectangular compound all four atoms have a short, medium, and long bond, ($s,m,l$).
In the triangular compound, two atoms (lower corners of the triangle) have the same distribution of bonds ($s,m,l$), but the lower middle and upper atoms have different distributions ($s,m,s$)  and ($l,m,l$). 
In \Ref{DistributionGraph2008} it has been shown that any simplex can be represented without loss of information, i.e., uniquely,
using such distributions of distances between vertices.

A continuous version of such a distribution of interatomic distances can be obtained by replacing
the scalar $\omega \times r_{IJ}$ argument in the cosine in the Fourier series by
an atomic radial distribution function $RDF_I$ for each atom $I$,
\bea
FDR(r) & = & \frac{1}{(2a)^{3}} e^{-\frac{r^2}{2a}} \sum_I Z_I^2 \cos[RDF_I(r)].
\label{eq:FDR}
\eea
In other words, the 1-D frequency domain $\omega$ has been turned into a 1-D real space interatomic distance domain.
Any functional form of atomic radial distribution functions, 
numerical or analytical such as ``softened'' Coulomb potentials $\sum_J Z_J/(|r-r_{IJ}|+1)$, or 
Slater (or Laplace) functions $\sum_J e^{-\alpha |r-r_{IJ}|}$ can be used.
Other smoothening functions, such as Gaussian radial distributions, were already proposed as descriptors in 
the past (See Refs.~\cite{Gasteiger1999,TodeschiniConsonniHandbookDescriptor,RDFdescriptor2006,bpkc2010,BartokGabor_Descriptors2013}).
To the best of our knowledge, however, they were not used as arguments in Fourier series expansions.

\begin{figure}
\centering
\includegraphics[scale=0.35, angle=0]{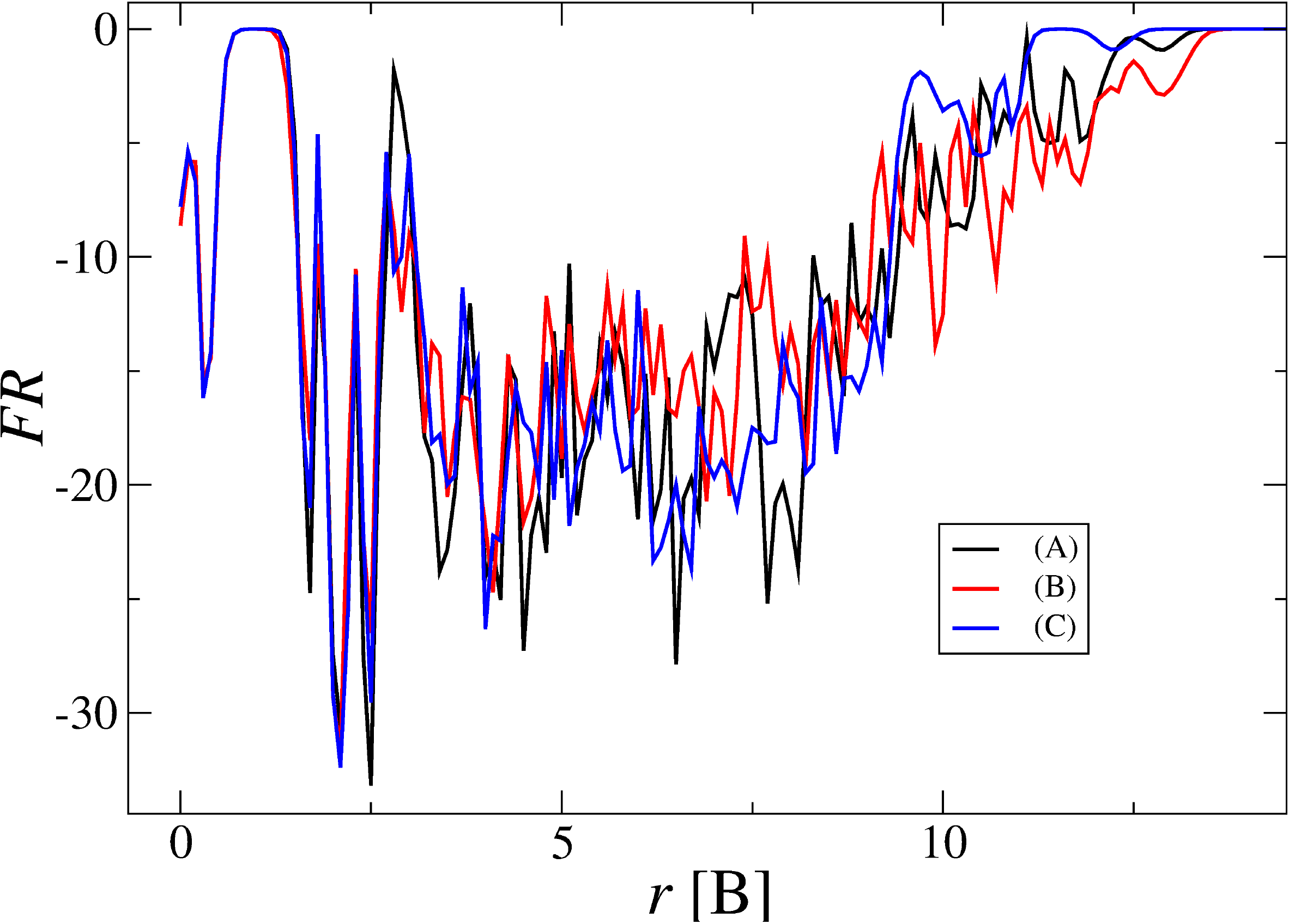}
\includegraphics[scale=0.07, angle=0]{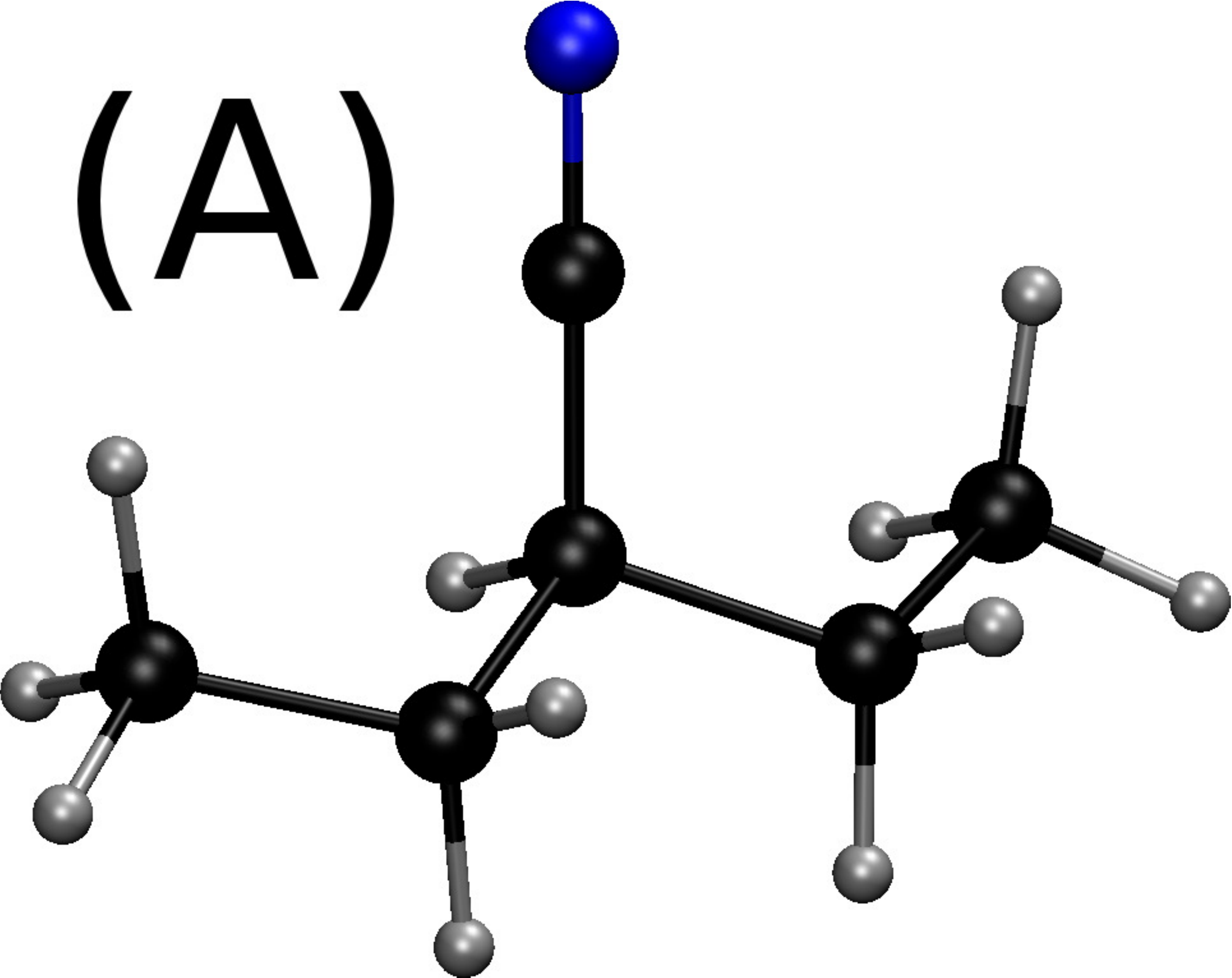}
\includegraphics[scale=0.07, angle=0]{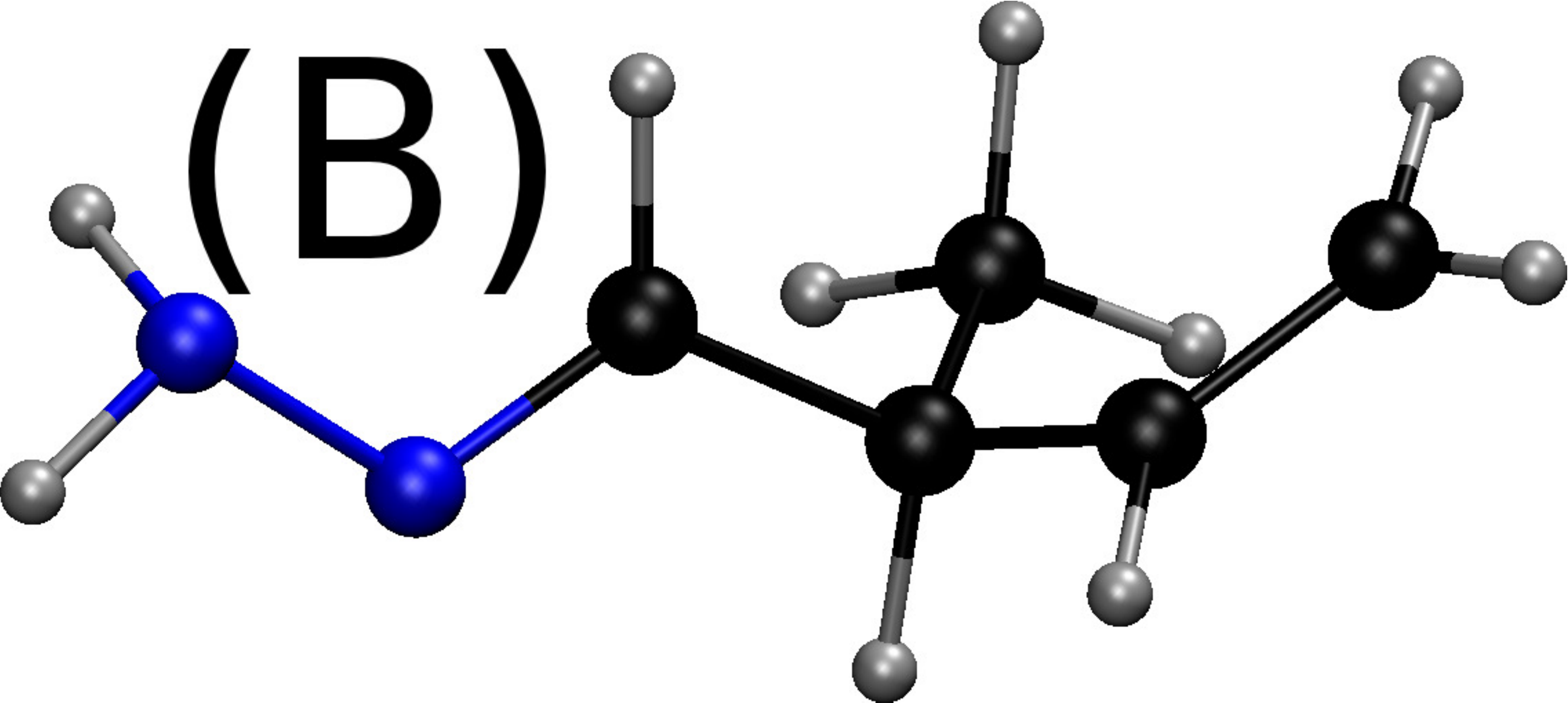}
\includegraphics[scale=0.07, angle=0]{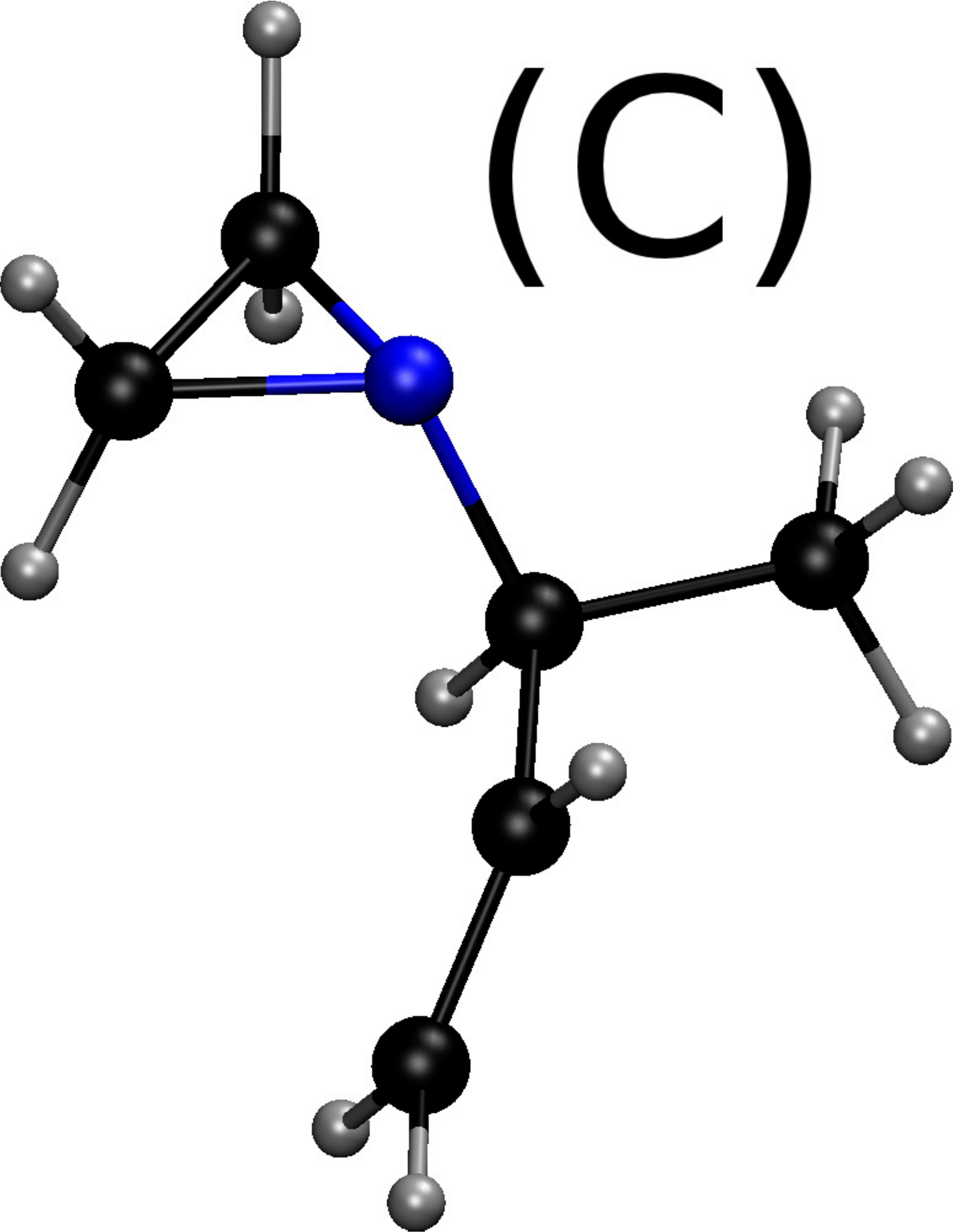}
\caption{
$FR$ fingerprints [Eq.~(\ref{eq:FR})] with optimized parameters $l, n, m, b$ (from 3k ML model) 
for three iso-electronic organic molecules (A), (B) and (C) with seven atoms
(not counting hydrogens), drawn at random from the GDB-7 data base~\cite{ReymondChemicalUniverse3}.
White, blue, and black atoms denote hydrogen, carbon, and nitrogen, respectively.
(A) and (C) are constitutional isomers while (B) differs in stoichiometry.
} \label{fig:FR3molecules}
\end{figure}

We have used ML models to test several variants of radial distribution
functions. The Gaussian radial distribution function,
$\sum_J Z_J^n e^{-b(r-r_{IJ})^2}/Z_I^m$, included
in the Fourier series in the following form, 
\bea
FR(r) & = & \sum_I  Z_I^l  \left( \cos\left[Z_I^m{\sum_J Z_J^n  e^{-b(r-r_{IJ})^2}}\right] - 1\right)
\nonumber \\
\label{eq:FR}
\eea
has resulted in the best performance. 
%f7 is variational in b and n_(i,Z). Here i \in {1,2,3}, n and b \in R, and b > 0
Here, we have omitted the Gaussian prefactor from Eq.~(\ref{eq:FDR}) to keep the descriptor complete, notwithstanding that it could still be used
to obtain finite length, or to localize the descriptor. 
Parameters $l, m, n, b \in \mathbb{R}$ are global hyperparameters which we optimize
via cross-validation when training the machine learning model. 
Further flexibility could still be introduced by making 
these parameters atom type $Z_I$-dependent. 
In this study, however, we have not investigated these degrees of freedom. 

$FR$ is a fingerprint as a function of interatomic distance. 
It decays to zero for all interatomic distances larger than the molecule.
The linear independence of the atomic terms in the Fourier summation, measurable by the Wronskian, guarantees that no atoms' RDF's linearly add (or cancel) each other---unless they {\em all} have exactly the same radial distribution. 
As such, the Fourier series introduces the linear independence of the radial distribution around each atom $I$. 
Only if all atoms in two molecules have the same $RDF$ will the two molecules yield the same $FR$ and therefore represent the same point on the potential energy surface.
Note that in \Ref{BartokGabor_Descriptors2013} an angular Fourier series-based descriptor that sums over individual angles (as opposed to distributions of angles) has been investigated. 
This descriptor, however, has been introduced in the context of modeling the potential energy surface of a single compound, not for training across CCS. 

The uniqueness of $FR$ can be recognized from a {\em Gedankenexperiment}: 
Imagine two $FR$s corresponding to two molecules.
In order for them to be the same, for each atomic term in the Fourier sum of one molecule there has to be an identical atomic term in the Fourier sum of the other molecule. 
This is only possible if the corresponding atoms in the respective molecules
happen to have the same $RDF_I$ (see below Eq.~(\ref{eq:FR}) for examples of atomic $RDF$s). 
Now, only if for each atomic $RDF_I$ in one molecule there is an identical atomic $RDF_I$ in the other molecule will the two $FR$ be the same, in which case the two molecules are identical (see also Boutin and Kemper~\cite{DistributionGraph2008}).

\section{Results}
%\begin{figure}[htbp]
%\centering
%\includegraphics[scale=0.3, angle=0]{plot_metricH2_H2.ps}
%\caption{
%Metric in chemical space, $D_{ij}$ (Eq.~(\ref{eq:Distance})), between two hydrogen %molecules
%$i$ and $j$ as a function of interatomic distance $r_{IJ}$ in one of them, 
%for 6 interatomic distances in the other. 
%Based on descriptor $FR$ (Eq.~(\ref{eq:FR}) with hyperparameter $b = 0.5$, 
%panel (b) is based on descriptor $FD_2$ (Eq.~(\ref{eq:FD_2}) with hyperparameter $a = 10$. 
%Note that as the interatomic distance of molecule $i$ approaches the one of molecule $j$
%the two molecules become identical, and consequently $D_{ij} \rightarrow 0$.
%Thus, the six lines correspond to the respective six interatomic distances for which $D_{ij}$ = 0.
%} \label{fig:Metric1}
%\end{figure}

\subsection{Organic molecules}
To illustrate the Fourier series of radial distribution function
descriptor for realistic systems, Fig.~(\ref{fig:FR3molecules}) features 
$FR$ for three iso-electronic organic molecules, drawn at random from the GDB data base~\cite{ReymondChemicalUniverse3}.
The nature of a molecular fingerprint, reminiscent of a spectrum, becomes evident for these more complex molecules.
Compound (B) has a different stoichiometry while (A) and (C) are 
constitutional isomers, differing merely in their covalent bonding pattern. 
Clearly, the fingerprints in Fig.~(\ref{fig:FR3molecules}) reflect the 
differences in molecular structure, in particular for larger distances. 
For smaller $r$, $FR$ can more easily be understood. 
For very small $r$ they look very similar, the first peak at $r < 0.5$ Bohr
is due to the stoichiometry (nuclear charges) only, with (B) being slightly off from
the $FR$ of (A) and (C) which are superimposed.
The subsequent three peaks (at 2 to 2.5 Bohr) reflect 
contributions from the first neighbor shell in the atomic radial distributions,
being also very similar, albeit not identical, for all three molecules with single, double, cyclic, or
even triple bonds (for (A)) between CH, NH, CC, CN, and NN atom pairs.
Note that the gap between stoichiometry and structural peaks in Fig.~(\ref{fig:FR3molecules})
can be expected to be conserved throughout CCS since there are no covalent bonds that are
shorter than bonds formed with hydrogen.
The fingerprints shown correspond to optimized hyperparameters settings in Eq.~(\ref{eq:FR}), 
alternative parameter combinations would lead to different appearance.

\subsection{Machine Learning models}
Any inductive approach requires us to measure distances in terms of input variables.
In order to compare chemical compounds, we use  
the Euclidean norm between the $FR$ descriptors of the two compounds as a proper metric.
More specifically, we consider the integral over the squared differences of two $FR$-descriptors
corresponding to molecules $i$ and $j$, 
\bea
\label{eq:Distance}
D_{ij}(r_{IJ}^{max})  & = & \sqrt{\int_{0}^{r^{max}_{IJ}} \;dr\; |FR_i(r) - FR_j(r)|^2}.  \nonumber
\eea
In our implementation, we discretized the $FR$ descriptor and employed an optimal value of
0.1 Bohr for the grid spacing, $dr$.
Note that for $\lim_{r_{IJ}^{max}\rightarrow \infty} D_{ij}$ converges for any molecular pair. 
Here, we used 20 Bohr as the integration upper bound, $r_{IJ}^{max}$, for all molecules.

\begin{figure}[htbp]
\centering
\includegraphics[scale=0.3, angle=0]{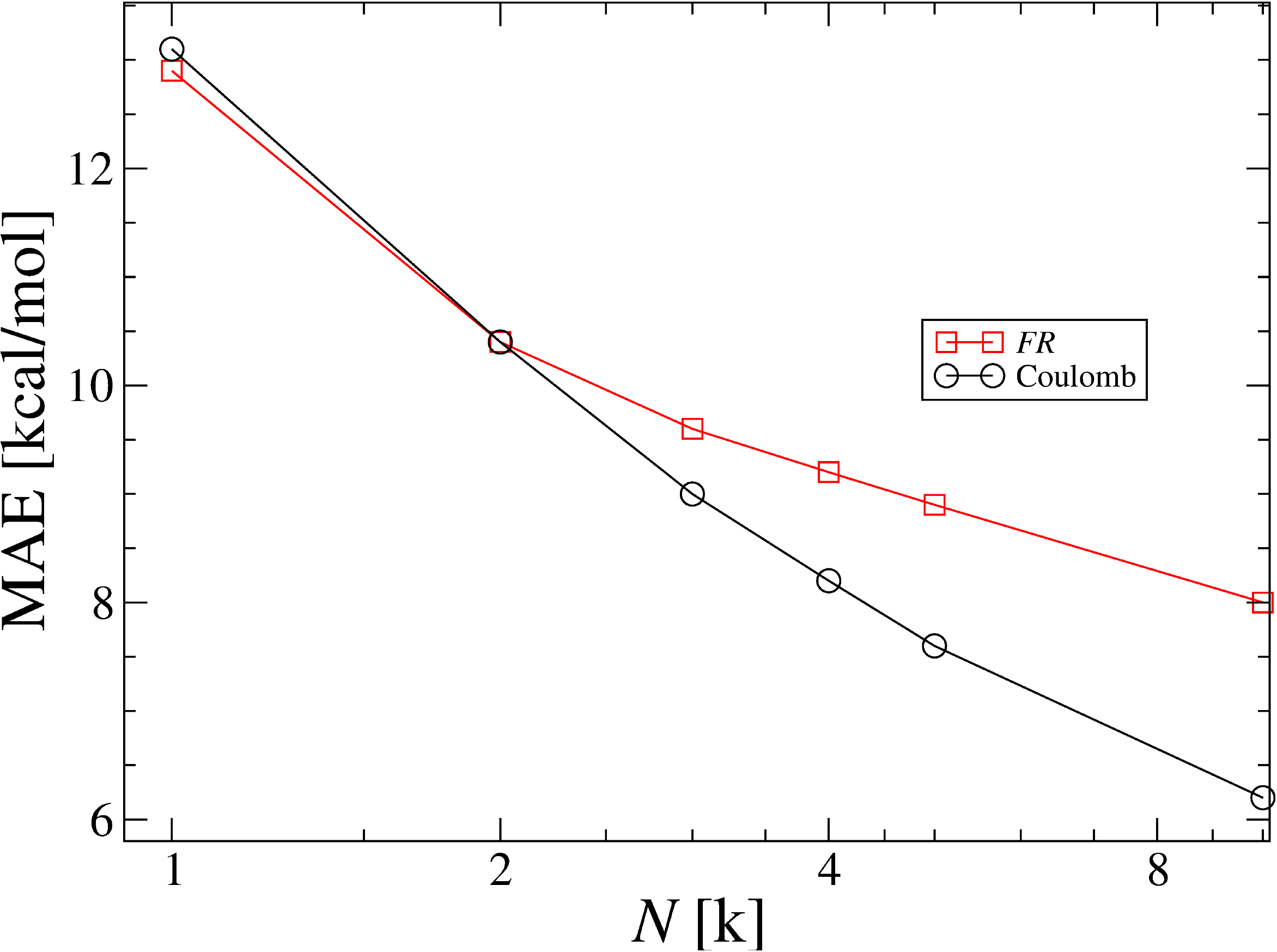}
\caption{
Mean absolute error (MAE) and root mean square error (RMSE) for out-of-sample predictions of atomization enthalpies at $T =$ 298.15 K,
as a function of training set size, $N$, for $FR$ and sorted $CM$ descriptor. 
Training and test sets consist of enthalpies of atomization at $T =$ 298.15 K
of the 134\,k molecules in the GDB-9 data base~\cite{ReymondChemicalUniverse3}, 
calculated at the B3LYP level of theory~\cite{DataPaper2014}. 
The inset shows the corresponding scatter plot for the 10k machine for predicted (ML)
versus actual (DFT) enthalpies of atomization in kcal/mol using sorted $CM$ (black) and $FR$ (red)
descriptors.
}
\label{fig:MLgdb}
\end{figure}

In order to have an idea of the $FR$'s performance of a descriptor, 
we have built ML models using the enthalpies of atomization for 134\,k organic molecules
taken from the GDB-17 database, recently published in \Ref{DataPaper2014}.
The GDB data base represents an exhaustive list of all organic molecules that can be constructed from up to 17 heavy atoms, containing as atom types C, N, O, F, S, Cl, Br, or I and saturating valencies with hydrogen atoms~\cite{ReymondChemicalUniverse,ReymondChemicalUniverse,ReymondChemicalUniverse2}.
All GDB molecules are expected to be stable and synthetically accessible according to organic chemistry rules~\cite{ReymondChemicalUniverse3}.
We have drawn at random training sets of sizes 1\,k, 2\,k, 3\,k, 4\,k, 5\,k, and 10\,k.

We then solved the kernel ridge regression problem for the given training sets following
the recipes set out in \Ref{AssessmentMLJCTC2013}. The solution yields
the coefficients $\{\alpha_i\}$ in the ML model of the atomization enthalpy $H$ 
with $FR$ as an input for any out of sample molecule $j$,
\bea
H(FR_j) & = & \sum_i^N \alpha_i k(D_{ij}).
\label{eq:KRR}
\eea 
Similarly, we have also trained ML models of the sorted Coulomb matrix $CM$
for the same training and testing sets. 
In the case of $FR$, a Gaussian kernel function $k$ with Euclidean
norm proved to lead to the best predictive performance. 
By contrast, in the case of $CM$ we used the Laplacian kernel function with
a Manhattan norm, following the findings in \Ref{AssessmentMLJCTC2013}.

Resulting mean absolute errors (MAE) and root mean square errors (RMSE) as a function of training set size $N$, as measured on the remaining molecules in the 134\,k set, is shown for both models in Fig.~\ref{fig:MLgdb}.
These error estimates have been obtained for out-of-sample predictions (not part of training set), 
using noise-level and length-scale hyper-parameters optimized through cross-validation 
runs on training sets. 
In the case of $FR$, also parameters $b, l, m, n$ in $FR$ [Eq.~(\ref{eq:FR})] have been optimized using
cross-validation for training set sizes $N$ = 1\,k, 2\,k and 3\,k. 
We have found that the training set size has relatively little influence
on these parameters, and we have therefore kept them fixed for all training set sizes larger than 3\,k. 
For the set of $N$ = 3\,k, optimal parameters $b, l, m, n$ amount to 7.0052, 0.0852, 1.2395, and -0.1626, respectively. Note for the construction of $FR$ that these parameters refer to interatomic distances in Bohr.
% paramB = 7.0051796460965194                                                         
%  nexp1 = 0.0851563347288825                                                             
%  nexp2 = 1.2394850956842507                                                             
%  nexp3 = -0.1625999325877687

The systematic decay of the mean absolute error with increasing training set size (see Fig.~\ref{fig:MLgdb}) 
is encouraging.
When compared to the current state of the art, the sorted Coulomb matrix, 
the $FR$ descriptor starts off at a slightly smaller MAE, and significantly smaller RMSE, for a training set size of 1\,k.
Up to 2\,k, $CM$ and $FR$ model errors decay with similar off-set and speed until they reach an
accuracy with MAE $\sim$11 kcal/mol,
an accuracy similar to the early generalized gradient approximated density 
functionals in Kohn-Sham DFT~\cite{ChemistsGuidetoDFT}. 
For larger training set sizes the MAE and RMSE of the $FR$ based model continue to decrease,
however at a decidedly slower learning rate. 
The $CM$-model's errors, in contrast, continue to decay significantly faster.
A possible explanation for the $FR$'s change in learning rate could be
that as the model's error passes 11 kcal/mol
remaining energy differences are dominated by differences in geometry
which, due to its high frequency oscillatory nature, the $FR$ descriptor possibly
captures only in a less efficient manner than the Coulomb matrix.
These first results, however, do not yet enable us to conclusively assess the $FR$'s performance.
Merely due to some inherent selection bias of the employed data sets 
the ML-model's performance using one descriptor might look favorable over the other. 
Here, for example, we used only relaxed geometries. 
When attempting to model reaction barriers, however,
the performance could possibly be inverted and lead to a different outcome. 
In any ways, the presented results do amount to numerical evidence suggesting
that for the modeling of atomization enthalpies further improvements are necessary before
the $FR$ descriptor can be considered competitive with the sorted $CM$ matrix.
Further improvements could possibly be achieved by making the $FR$ 
hyperparameters atom type $Z_I$-dependent. 

\subsection{Computational details}
Hyperparameters were estimated through 5-fold cross validation (CV) on training set of size $N$.
Accordingly, $N$ training molecules were distributed at random into 5 bins, each containing $N$/5 molecules.
Each bin was used once as the holdout set, with the remaining 4 bins as training set, and hyperparameters were optimized by minimizing the MAE for the holdout-bin.
Globally optimal hyperparameters were obtained by taking the median of the 5 folds.
The final kernel with globally optimized hyperparameters was subsequently used to predict
atomization enthalpies for the 134\,k$-N$ out-of-sample molecules which never had a part in training.

\section{Conclusions}
\label{sec:Discussions}
A set of fundamental physical arguments has been introduced as to what are 
crucial and desirable properties of descriptors that can be expected to yield reliable 
performance in intelligent data analysis (IDA) methods when applied to the modeling
of quantum chemical properties of molecules. 
Starting from the external potential in the electronic Hamiltonian, and 
using Fourier transforms and radial distribution functions, 
we have introduced an intramolecular distance based fingerprint-like descriptor, $FR$, corresponding to 
a Fourier series of atomic radial distribution functions.
The $FR$ is unique for any molecular compound (i.e. chemical composition and geometry), 
and invariant with respect to translation, rotation, and atom indexing.
Furthermore, $FR$ is differentiable, not only with respect to nuclear displacement for geometry optimization or molecular dynamics, but also with respect to ``alchemical'' changes, i.e. change in nuclear 
charges~\cite{AlchemicalDerivativeBinaryMetalCluster_WeigendSchrodtAhlrichs2004, anatole-jcp2006-2, anatole-jcp2007, anatole-jctc2007-2, anatole-jcp2009-2}, 
potentially useful for computational materials design~\cite{anatole-prl2005,CatalystSheppard2010,anatole-ijqc2013}.
As such, this descriptor exhibits all the crucial and desired properties listed in Table~\ref{tab:Properties}.
The $FR$ descriptor can be reduced to $N\times(N-1)/2$ dimensionality if it is evaluated only at those $r$-values that correspond to interatomic distances in a compound. 
A Gaussian pre-factor can be used to damp the descriptor to reduce the dimensionality further and to introduce locality. 
Results from preliminary ML models, yielding promising predictive power for out-of-sample compounds, 
suggest that the $FR$ descriptor, or variants thereof, is likely to be well suited for the generic and systematic construction of ML models that are valid for all regions of the potential energy surface 
of novel compounds, as long as trained across sufficiently representative subspace of CCS.  
The current $FR$-performance, however, is not (yet) on par with the sorted Coulomb matrix.
Note that the $FR$ exclusively represents the external potential of a molecule, not the molecule's charge. Differences in the latter can easily be added to $FR$ distances through the use of more sophisticated metrics, 
such as normalized Euclidean, or Mahalanobis, distances. 
A more comprehensive assessment, also including non-equilibrium geometries on the same potential energy surfaces, will be subject of future work. 

\section{Acknowledgments}
The authors thank J.~R.~Hammond, M.~Hereld, and A.~Vazquez-Mayagoitia for discussions and suggestions. 
This research used resources of the Argonne Leadership Computing Facility at Argonne National Laboratory, 
which is supported by the Office of Science of the U.S. DOE under Contract No. DE-AC02-06CH11357.
OAvL acknowledges support from LDRD funding (Multiscale Materials Modeling using Accurate Ab Initio Approaches (M3A3)).
OAvL acknowledges funding from the Swiss National Science foundation No. PP00P2\_138932.
Some of the calculations were performed at sciCORE (http://scicore.unibas.ch/) scientific computing core facility at University of Basel.
\bibliography{literatur}
\end{document}